\documentclass[9pt,twocolumn,twoside]{osajnl}
\usepackage{gensymb}
\usepackage{soul}
\usepackage{tabularx}
\journal{ao} 
\usepackage[justification=justified]{caption}
\usepackage{amsmath}

\setboolean{shortarticle}{false}
\newcommand{\dg}{$^{\circ}$} 

\title{The Simons Observatory: Metamaterial Microwave Absorber (MMA) and its Cryogenic Applications}

\author[1, 2]{Zhilei Xu}
\author[3]{Grace E.~Chesmore}
\author[4]{Shunsuke Adachi}
\author[5]{Aamir M.~Ali}
\author[6]{Andrew Bazarko}
\author[1, 7]{Gabriele Coppi}
\author[1]{Mark Devlin}
\author[8]{Tom Devlin}
\author[1]{Simon R.~Dicker}
\author[9]{Patricio A.~Gallardo}
\author[3]{Joseph E.~Golec}
\author[10]{Jon E.~Gudmundsson}
\author[11]{Kathleen Harrington}
\author[12]{Makoto Hattori}
\author[1]{Anna Kofman}
\author[13]{Kenji Kiuchi}
\author[14, 12, 15, 16]{Akito Kusaka}
\author[1]{Michele Limon}
\author[17]{Frederick Matsuda}
\author[11, 18, 19, 3]{Jeff McMahon}
\author[7]{Federico Nati}
\author[9, 20]{Michael D.~Niemack}
\author[3]{Shreya Sutariya}
\author[14]{Aritoki Suzuki}
\author[21]{Grant P.~Teply}
\author[22, 1]{Robert J.~Thornton}
\author[23]{Edward J.~Wollack}
\author[7]{Mario Zannoni}
\author[1]{Ningfeng Zhu}

\affil[1]{Department of Physics and Astronomy, University of Pennsylvania, 209 South 33rd Street, Philadelphia, PA 19104, USA}
\affil[2]{MIT Kavli Institute, Massachusetts Institute of Technology, Cambridge, MA 02139, USA}
\affil[3]{Department of Physics, University of Chicago, 5720 South Ellis Avenue, Chicago, IL 60637, USA}
\affil[4]{Division of Physics and Astronomy, Graduate School of Science, Kyoto University, Kitashirakawa-Oiwakecho, Sakyo-ku, Kyoto 606-8502, Japan}
\affil[5]{Department of Physics, University of California - Berkeley, Berkeley, CA, USA}
\affil[6]{Department of Physics, Princeton University, Princeton, NJ, USA}
\affil[7]{Department of Physics, University of Milano-Bicocca Piazza della Scienza, 3 - 20126 Milano (MI), Italy}
\affil[8]{Devlin Design Inc.}
\affil[9]{Department of Physics, Cornell University, Ithaca, NY 14853, USA }
\affil[10]{The Oskar Klein Centre, Department of Physics, Stockholm University, AlbaNova, SE-10691 Stockholm, Sweden}
\affil[11]{Department of Astronomy and Astrophysics, University of Chicago, 5640 S. Ellis Ave., Chicago, IL 60637, USA}
\affil[12]{Astronomical Institute, Graduate School of Science, Tohoku University, 6-3, Aramaki Aza-Aoba, Aoba-ku, Sendai 980-8578, Japan}
\affil[13]{Department of Physics, The University of Tokyo, 7-3-1 Hongo, Bunkyo-ku, Tokyo 113-0033, Japan}
\affil[14]{Physics Division, Lawrence Berkeley National Laboratory, 1 Cyclotron Road, Berkeley, CA 94720, USA}
\affil[15]{Kavli Institute for the Physics and Mathematics of the Universe (WPI), Berkeley Satellite, the University of California, Berkeley 94720, USA}
\affil[16]{Research Center for the Early Universe, School of Science, The University of Tokyo, Tokyo 113-0033, Japan}
\affil[17]{Kavli Institute for the Physics and Mathematics of the Universe (WPI), The University of Tokyo Institutes for Advanced Study, The University of Tokyo, Kashiwa, Chiba 277-8583, Japan}
\affil[18]{Kavli Institute for Cosmological Physics, University of Chicago, 5640 S. Ellis Ave., Chicago, IL 60637, USA}
\affil[19]{Enrico Fermi Institute, University of Chicago, Chicago, IL 60637, USA }

\affil[20]{Department of Astronomy, Cornell University, Ithaca, NY 14853, USA}
\affil[21]{Department of Physics, University of California, San Diego, CA 92093-0424, USA}
\affil[22]{Department of Physics and Engineering, West Chester University of Pennsylvania, West Chester, PA, 19383, USA}
\affil[23]{Goddard Space Flight Center, 8800 Greenbelt Road, Greenbelt, MD 20771, USA}

\affil[*]{Corresponding author: \href{mailto:zhileixu@sas.upenn.edu}{zhileixu@sas.upenn.edu}}

\begin{abstract}
Controlling stray light at millimeter wavelengths requires special optical design and selection of absorptive materials, which should be compatible with cryogenic operating environments.  While a wide selection of absorptive materials exist, these typically exhibit high indices of refraction, which reflect/scatter a significant fraction of light before absorption. For many lower index materials such as commercial microwave absorbers, their applications in cryogenic environments are challenging. 
In this paper we present a new tool to control stray light: metamaterial microwave absorber (MMA) tiles. These tiles are comprised of an outer metamaterial layer, which approximates a lossy gradient index anti-reflection coating. They are fabricated via injection molding commercially-available carbon-loaded polyurethane (25\% by mass). The injection molding technology enables mass production at low cost.  The design of these tiles is presented, along with thermal tests to 1\,K. Room temperature optical measurements verify their control of reflectance to less than 1\% up to 65$^{\circ}$ angles of incidence, and control of wide angle scattering below 0.01\%. The dielectric properties of the bulk carbon-loaded material used in the tiles are also measured at different temperatures, confirming that the material maintains similar dielectric properties down to 3\,K.
\end{abstract}

\setboolean{displaycopyright}{true}

\begin{document}

\maketitle

\section{Introduction}
The Simons Observatory (SO) is a series of millimeter-wave telescopes designed to observe the Cosmic Microwave Background (CMB) temperature and polarization signals to an unprecedented sensitivity~\cite{gali18, so19}. With the combination of one Large Aperture Telescope (LAT)~\cite{xu/etal:2020c, zhu18, orlo18, coppi/etal:2018} and three Small Aperture Telescopes (SAT)~\cite{ali20}, the experiment will measure the temperature and polarization anisotropy of the cosmic microwave background with $\sim$\,70,000 background noise limited detectors operating at $\sim$\,100\,mK. SO will test cosmic inflation during the early universe, characterize the primordial perturbations, measure the effective number of relativistic species and the sum of the neutrino masses, and improve our understanding of galaxy evolution and the era of cosmic reionization~\citep{so19}. 

Modern ground-based millimeter-wave telescope receivers have advanced to the point where detector noise is dominated by photon noise, meaning the thermal or electronic noise intrinsic to the detectors (and the associated readout system) is less than the photon noise from the incident light. The in-band optical power incident on the detectors---closely related to the photon noise---arises from the sky signal, the atmosphere, and stray light outside of the desired optical path. The out-of-band instrument spectral and angle response is largely set by filter stack and baffle implementation.
The sky signal and the atmosphere power loading observed by the detectors cannot be reduced by improving the instrument, whereas the stray light loading can be reduced by a more effective design. Instead of following the main optical path, stray light is reflected or scattered by the side walls of cryogenic receivers before being absorbed by the detectors. Therefore, stray light can not only compromise image fidelity through ghosting or glint, but also degrades the detector sensitivity. In this paper we concentrate on suppressing stray light by minimizing reflection and scattering within cryogenic receivers~\cite{iuliano/etal:2018, thornton/etal:2016, sharp/etal:2008}.

Terminating stray light at the lowest possible temperature is typically achieved by cryogenic baffling design which minimizes reflection and scattering.  Ideally, the baffling surfaces are constructed of millimeter-wave absorbing material, with high absorptivity and low reflection/scattering over a wide range of incident angles and frequencies. Beyond the required optical properties, the absorbers should be efficiently cooled to cryogenic temperatures with robust thermal contacts and be able to withstand the mechanical stress associated with thermal cycling. Since cryogenic space is scarce, the absorbing material is often required to be as compact as possible. Ideally, the absorber should be light in mass, especially for future space missions.

Available options include conductively-loaded-epoxy~\citep{Wollack2008}, and commercially-available absorptive sheets\footnote{For example, HR-10 sheets from Emerson\,\&\,Cummings.}/tiles\footnote{For example, Tessellating TeraHertz Radar Absorbing Materials (RAM), Thomas Keating Ltd.}. The conductively-loaded-epoxy relies on the conductive materials to absorb the radiation; however, the high index of refraction of the epoxy gives high reflectance and scattering, depending on the roughness of the surface. In addition, the epoxy normally has a $>$\,2\,g$\cdot$cm$^3$ density, adding a significant mass to the cryogenic system. The commercially-available absorptive sheets and tiles provide high absorption, while mechanically and cryogenically attaching them to low-temperature surfaces poses great challenges. Recent development on 3D printing also enables more absorber designs~\cite{petroff/etal:2019}, while mass production at low cost is still a challenge. To overcome these obstacles, we developed injection-molded tiles with a metamaterial gradient index anti-reflection coating. Metamaterials are artificially engineered materials, in terms of content and geometry, to achieve physical properties that are not available in nature~\cite{wollack/etal:2016, ding/etal:2012, watts/liu/padilla:2012}. Our solution provides high absorption with customized thermal and mechanical interface for cryogenic applications. 

We describe the design (Section~\ref{sec:optical_design},\,\ref{sec:mechanical_design}) and characterization (Section~\ref{sec:thermal_testing},\,\ref{sec:optical_testing}) of the tiles, along with its application within the Simons Observatory (SO) Large Aperture Telescope Receiver (LATR)~\cite{zhu18, orlo18} and Small Aperture Telescopes (SAT)~\cite{ali20}.\footnote{The MMA tiles are also used in CCAT Prime-cam~\cite{vavgiakis/etal:2018}.} We also discuss potential future applications in Section~\ref{sec:future_applications}.

\section{Optical Design}
\label{sec:optical_design}

\begin{figure}[t]
    \centering
    \includegraphics[width = .45\textwidth]{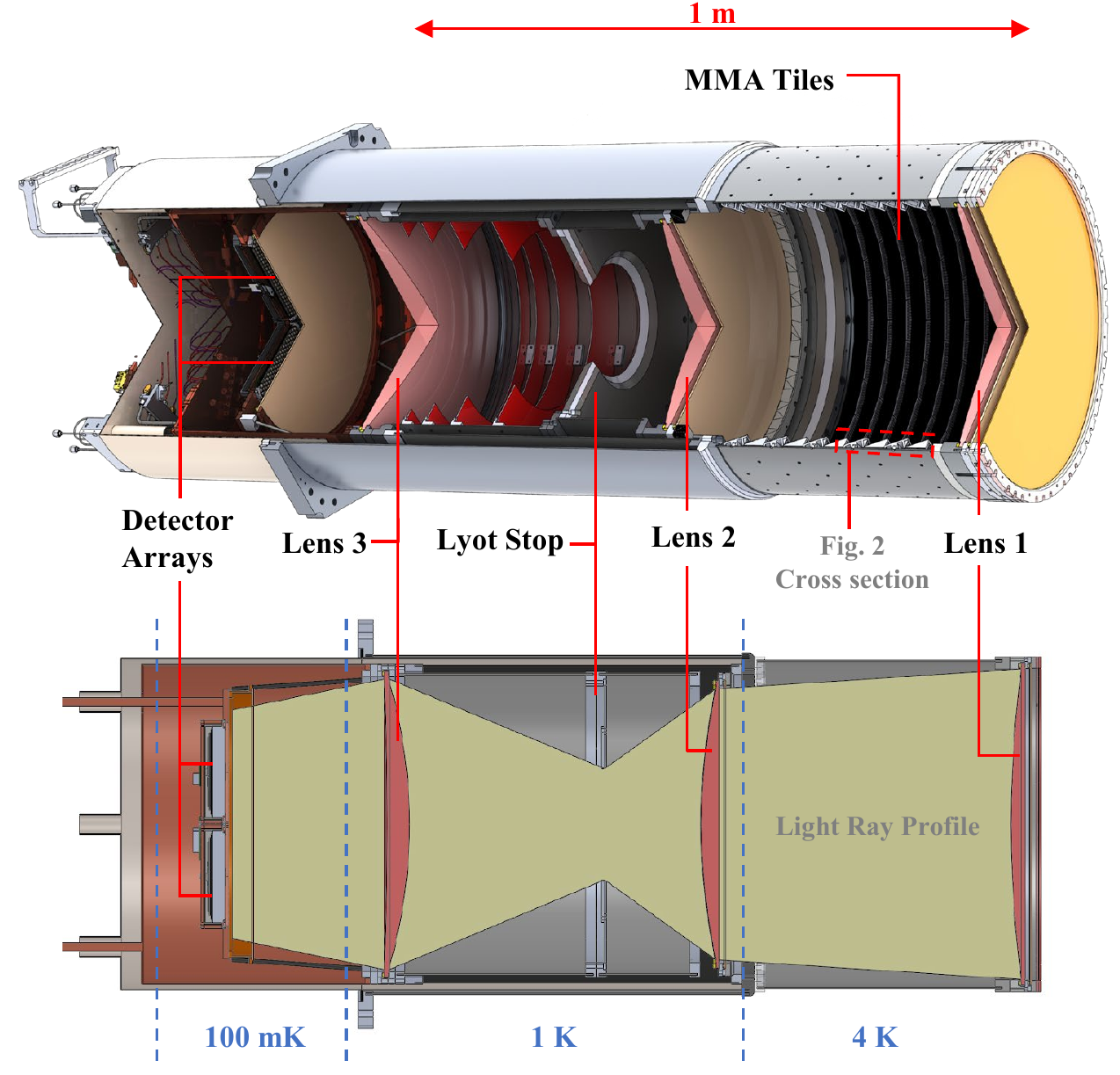}
    \caption{Large Aperture Telescope Receiver (LATR) optics tube in the Simons Observatory. A detailed rendering of the optics tube design is shown in the upper part and a simplified schematic cross section is shown in the bottom part. Incoming light enters the optics tube from the right and focuses on the detector arrays on the left via three lenses. Yellow-shaded region in the bottom part shows the geometric light ray profile. Operational temperatures for different parts of the optics tube are also annotated. More details of the optics tubes are described in~\cite{xu/etal:2020c}. The MMA tiles are installed in the regions between lens\,1 and lens\,2 at 4\,K. Fig.~\ref{fig:tile_design} shows a cross section of the MMA tile assembly (box with red dashed lines) in the optics tube and the detailed design of one tile. A flat version of the MMA tile (Section~\ref{sec:future_applications}) was also used to cover the front and back of the Lyot stop at 1\,K.}
    \label{fig:latr_ot}
\end{figure}

\begin{figure*}[t]
    \centering
    \includegraphics[width = .9\textwidth]{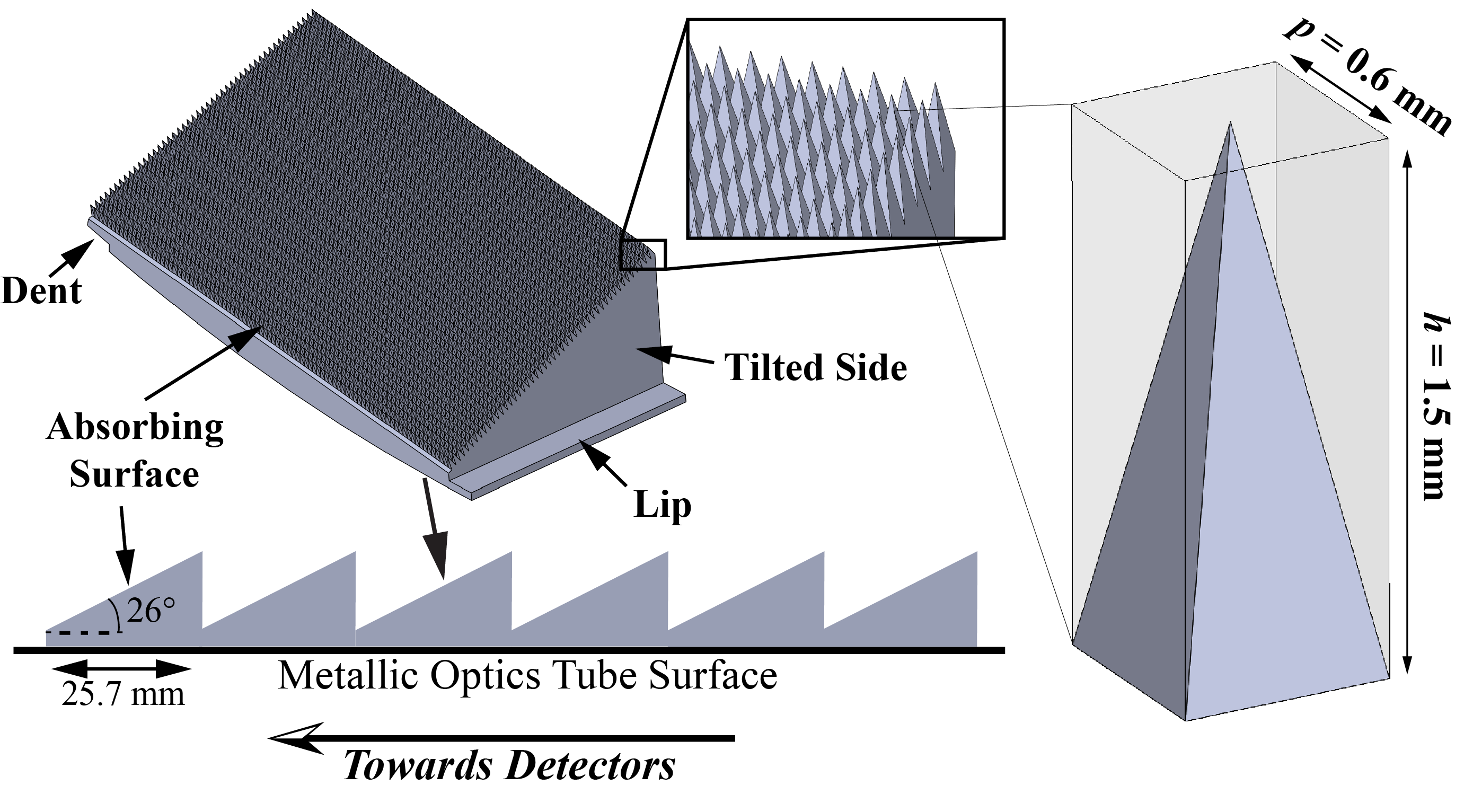}
    \caption{MMA tile design. A snapshot of a single tile design is shown. The tilted upper surface has the pyramidal structures as described in Section~\ref{sec:optical_design}. A zoom-in plot shows details of the pyramidal structures, acting as a metamaterial anti-reflection coating. Another insert on the right zooms in further and shows one single unit of the metamaterial structure, with a pitch, $p =0.6\,$mm and a height, $h = 1.5\,$mm. The tile bottom is curved to fit the cylindrical inside surface of the metallic optics tube, which is used for structural support, heat sinking, and reflective optical termination for the absorptive tiles. A lip and a dent are for seamless tessellating. A cross section of the segmented tilted surface is shown in the lower left part. In a time-reverse fashion, light rays coming from the left would hit the tilted surface with a <\,64\dg{} angle of incidence, where the absorbing surface achieves the desired optical performance.}
    \label{fig:tile_design}
\end{figure*}

The ideal baffling material minimizes reflection and scattering for light over the relevant range of angles of incidence, while mechanically (and even cryogenically) fitting within optical systems.  In the case of the SO LATR~\cite{xu/etal:2020c, zhu18, orlo18, coppi/etal:2018}, the desire to close-pack the optics tubes limits the radial extent of the absorbing material to about 1\,cm to avoid clipping the beam near the front of the optics tubes (between Lens\,1 and Lens\,2 in Fig.~\ref{fig:latr_ot}). The tight radial space prohibits the application of ring baffles~\cite{iuliano/etal:2018, thornton/etal:2016, sharp/etal:2008}, which are implemented in the middle section of the LATR optics tubes (between Lyot stop and Lens\,3 in Fig.~\ref{fig:latr_ot}) and the SO small aperture telescopes (SAT)~\cite{ali20}. 

A detailed study of the SO LATR optics tubes revealed that improving light absorption near the front of the optics tubes would most significantly reduce optical power loading~\cite{gudmundsson/etal:2020}. This study also showed that stray light on that section of the optics tubes ranges from $55^\circ$ to $90^\circ$ relative to the surface normal, taking up $\sim$\,2\,\% of the solid angle. Controlling the stray light in that section improves the telescope's mapping speed by 40\,--\,80\%,\footnote{Mapping speed is defined as $1/\textrm{NET}^2$, where NET is the noise-equivalent-temperature (NET). More details about mapping speed are available in \cite{hill/etal:2018}.} significantly increasing the sensitivity of the instrument. The Fresnel Equations show that the reflectance for dielectric materials approaches unity near grazing angles.  To improve performance at these high angles the absorbers were fabricated with the absorbing surface tilted at $26^\circ$ relative to the normal of the optics tube surface (Fig.~\ref{fig:tile_design}). The tilting reduces the angle of incidence by 26\dg{} to $<64^\circ$ (incoming light cannot exceed 90\dg{} angle of incidence) where the absorbing surface provides desired optical performance. Note that the angle of incidence is viewed in a time-reverse fashion here.

The goal of mass production led to the selection of injection molding carbon-loaded plastics. The carbon content increases the optical loss of the plastic materials so that the radiation is sufficiently attenuated within a small depth of bulk materials. However, the carbon-loaded plastics have relatively high dielectric functions, leading to high reflectance on a flat surface. Therefore, an effective anti-reflection coating is designed.

At the start of the design process, the exact material properties were not known. Furthermore, it was anticipated that variations in the composition, preparation, and processing of the material could lead to significant variations in the dielectric properties of the final materials. In addition, manufactured structures may not realize the design geometry perfectly. For these reasons, the design is based on a gradient index comprised of sub-wavelength pyramids with a height-to-pitch ratio as 2.5 ($h/p = 2.5$), which corresponds to a pyramid vertex half-angle of $\sim$\,11\dg{} (see Fig.~\ref{fig:tile_design}). The design is similar to that shown in \cite{Chuss2017} (primarily Fig.~1 and Fig.~2 in that paper). For wavelengths shorter than twice of the pitch, diffractive scattering rises; for wavelengths longer than height of the pyramids, the reflectance approaches the bulk dielectric material and the AR coating is ineffective~\cite{wollack/etal:2016}. Meanwhile, the $h/p$ ratio is limited by the plastic mechanical properties. The overall design utilizes the metamaterial concept at two levels: the bulk material is a mixture of plastic loaded with carbon particles to achieve high absorption; furthermore, the graded index metamaterial is tiled on the absorbing surface. This layer serves as an adiabatic impedance matching structure between free space and the bulk conductively loaded dielectric.

\begin{figure*}
    \centering
    \includegraphics[width = .9\textwidth]{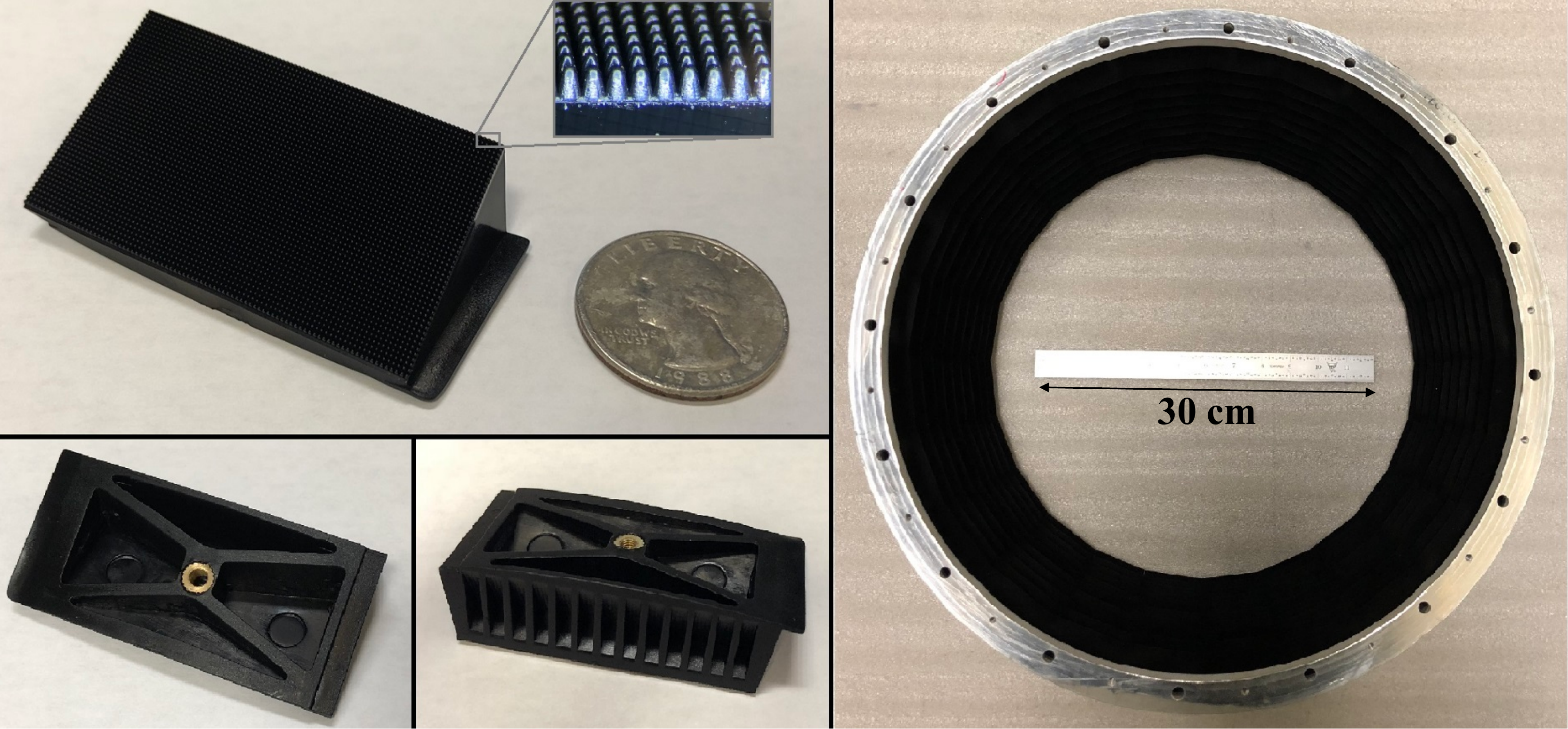}
    \caption{Manufactured MMA tiles. The manufactured tiles are presented from different perspectives. The top left photo shows one tile from the absorbing surface. A microscope photo of the pyramids is shown in the insert. The pitch between pyramids is kept as 0.6\,mm as designed. The bottom left two photos show the back of the tile from two other perspectives. The lightweighting pockets are visible along with the brass injected M3 nut for fastening. The photo on the right shows the assembly of 240 tiles installed on the wall of the optics tube section (see Fig.~\ref{fig:latr_ot}). A ruler ($\sim$\,30\,cm) is placed in the center for scale. Note that the absorbing surface appears to be black and featureless in the assembly photo. The condition stays the same regardless of lighting in the room, which supports the effectiveness of the AR-coating design even in optical.}
    \label{fig:real_tile}
\end{figure*}

The optical performance was simulated by High Frequency Structure Simulator (HFSS).\footnote{HFSS: https://www.ansys.com/products/electronics/ansys-hfss} The initial simulations, assumed a relative dielectric permittivity of $\hat{\varepsilon_{\textrm{r}}} \approx 4 + 0.66i$ and ideal geometry, predicted reflectance below $1\%$  (5\%) for angles of incidence up to $45^\circ$ ($60^\circ$) above 75\,GHz. In comparison, simulations using the manufactured geometry (see Fig.~\ref{fig:real_tile}) and measured dielectric function of the bulk material at $\sim$\,40\,GHz (Equation~\ref{equ:dielectric_function}) predict reflectance below $2\%$  (8\%) for the same angles of incidence and frequency range.  This performance changed within a factor of two to the initial simulation even though the input material dielectric function changed by a factor of four and the pyramid shapes are different. However, the simulated reflectance from the model are higher than the measured performance.   
It is noteworthy that the carbon particle size (50--75\,$\mu$m) is not insignificant compared to the manufactured tips of the pyramids ($\sim$\,150\,$\mu m$ in the product, Fig.~\ref{fig:real_tile}), suggesting an uneven carbon particle filling from the bottom to the top of the pyramids. Therefore, stratification of the carbon loading near the tips is likely during the molding process.  
This implies there can be a gradient in the dielectric properties. To explore this, a model with measured pyramidal features and a four-layer approximation to a gradient in the dielectric properties was constructed.   This reduced reflectance and introduced a frequency dependence similar to what is observed in Section~\ref{sec:optical_testing}. The simulations do not indicate scattering below the frequency of 250\,GHz. At the frequency of 300\,GHz, the simulations indicate that scattering appears to be at most percent level. 
This family of simulations highlights the robustness of this design to changes in the dielectric properties and to a lesser extent the geometry, which is the key feature of this design.

In developing the fabrication process, several carbon loaded material formulations were explored (e.g., carbon loaded polypropylene, carbon loaded polyethylene, carbon-fiber loaded polypropylene, and carbon loaded polyurethane). Considering manufacturability and cryogenic robustness (more details in Section~\ref{sec:thermal_testing}), we chose thermoplastic polyurethane (TPU) with 25\% carbon loading by mass.\footnote{TPU materials with >\,25\% carbon content tend to fracture.} The loaded carbon content is conductive carbon powder with mesh between 200\,--\,300, corresponding to 50\,--\,75\,$\mu$m in diameter.\footnote{The selected material, \href{http://www.qianyan.biz/qy/item.do?i=4&page_no=2&yid=10876405&sid=1657904}{Conductive TPU Compound}, is a commercially-available electromagnetic interference shielding material from Yushuo New Material Ltd.}

Several waveguide samples with the bulk TPU material filling the 2:1 rectangular cross section were prepared in WR-28 (i.e., guide broadwall 7.11 mm and shim thicknesses of 1.4 and 5.6 mm) and WR-22 (i.e., guide broadwall 5.69 mm and shim thickness 4.8 mm). Two different methods were used to realize each waveguide sample type: (1) press fitting machined dielectric samples into a split-block waveguide fixture and (2) directly thermally molding the material into a waveguide shim. Then the complex dielectric function,
\begin{equation}
\label{equ:dielectric_function}
\hat{\varepsilon_{\text{r}}} = \varepsilon_r' + \varepsilon_r''\,i \approx 16 + 11\,i,
\end{equation}
was measured by a vector network analyzer in the frequency range from 26\,GHz to 52\,GHz~\cite{Wollack2008}. All of the samples had the same real part of the dielectric function. The imaginary parts were similar for the cast samples, but were found to be 25\% higher for the mechanically inserted piece.  We believe that this could be due to the presence of voids in the cast samples, mechanical stress in the inserted sample or some combination.  This implies the dielectric function of this material has some degree of process dependence. Another transmission measurement implies the dielectric function changes <\,5\,\% for the real part and <\,50\,\% for the imaginary part from 20\,GHz to 200\,GHz. However, the metamaterial layer structure is designed to be robust to the variation in the complex dielectric function, which is verified by the optical measurements in Section~\ref{sec:optical_testing}. 

Given the measured dielectric properties, the field penetration depth is calculated as, 
\begin{equation}
\delta = \frac{\lambda}{2\pi\kappa} \simeq 0.12\, \lambda,
\end{equation}
where 
\begin{equation}
\kappa = {1 \over \sqrt{2}}  \sqrt{ \left( \varepsilon_r'^2 + \varepsilon_r''^2 \right)^{1 \over 2}  - \varepsilon_r'  } \simeq 1.3
\end{equation}
is the imaginary component of the refractive index. $\varepsilon_r'$ and $\varepsilon_r''$ are the real and imaginary part of the dielectric function in Equation~\ref{equ:dielectric_function}.  For a free space radiation with a wavelength of $\lambda$, one obtains a power attenuation, 
\begin{equation}
    P/P_o = e^{-2\, z/\delta},
    \label{equ:attenuation}
\end{equation}
showing that the radiation is attenuated by a factor of $1/e$, every $\delta/2$ of the bulk thickness $z$. This equation defines the designed bulk material thickness to achieve the desired low-transmission from the tiles.

\section{Mechanical Design and Fabrication}
\label{sec:mechanical_design}

A cross section of the segmented surface is shown in the lower part of Fig.~\ref{fig:tile_design}, which is along the optics tube axis. Each of the segments forms a ring rotating around the cylinder axis. Within one ring, we also separate the structure into 24 arcs, each spanning 15\dg{}. In this way, the cylindrical surface is decomposed into one single tile.\footnote{In practice, the tile is designed to occupy 99\% of the allocated dimensions, say spanning 14.85\dg{} instead of 15\dg{}, to account for manufacturing tolerance.}

A snapshot of the tile design is presented in the upper left part of Fig.~\ref{fig:tile_design}. The absorbing surface is covered with pyramidal structures, acting as an impedance-matching anti-reflective coating. Beneath the coating, the design has at least 2\,mm of bulk material to absorb the radiation. According to Equation~\ref{equ:attenuation}, radiation at different wavelengths will be effectively attenuated by
\begin{align}
    P/P_o  \simeq 2.2\times10^{-10} &\simeq -96\ \textrm{dB}\ (\lambda = 1.5\,mm),\\
    P/P_o  \simeq 1.5\times10^{-5} &\simeq  -48\ \textrm{dB}\ (\lambda = 3\,mm).
    \label{equ:attenuation_calc}
\end{align}
These tiles are installed on metal surfaces and waves within the bulk absorber layer experience at least twice the calculated attenuation factor before leaving this volume. Adjacent to the absorbing surface, two sides are also designed to tilt inwards to properly form a ring once assembled. A lip and a dent are designed for tessellating to eliminate gaps between the tiles in an assembly. The bottom of the tile is curved to match the surface of the optics tube cylindrical wall.

Fastening hundreds of the tiles in a metal cylinder raises another challenge. Ideally, there should be only one point of contact for one tile: at cryogenic temperatures, differential thermal contraction between the metal mounting structure and the plastic tiles could cause a stress failure if the tile is constrained at more than one point. On the other hand, one mechanical contact limits the thermal conductance for cooling the tile. Therefore, each tile is only $3$\,$\times$\,5\,cm in size, small enough that one contact is sufficient to conduct heat away at cryogenic temperatures. This is also a convenient size for mass production.

During injection molding process, a metal mold is first machined to high precision. Then the melted plastic is injected into the mold at high pressure to form the designed shape before it is removed as one piece. 
The injection molding process employed can achieve a minimum feature size of $\sim$\,100\,$\mu$m and the release process is compatible with geometries having a monotonically increasing surface profile.\footnote{This constraint can be relaxed for an additive manufacturing process (e.g. 3D printing, as published in~\cite{petroff/etal:2019}).}
After the injection molding process, a brass M3 nut is installed by heat-pressing as the single point of fastening. There are several lightweighting features on the back and the bottom of the tile. Those features were developed in collaboration with an injection molding manufacturer to help the tile maintain its shape during manufacture. They are also designed with the necessary drafting angles to release the part from the injection mold. Once the mold is available, thousands of the tiles are manufactured in a month.

\begin{figure}
    \centering
    \includegraphics[width = .45\textwidth]{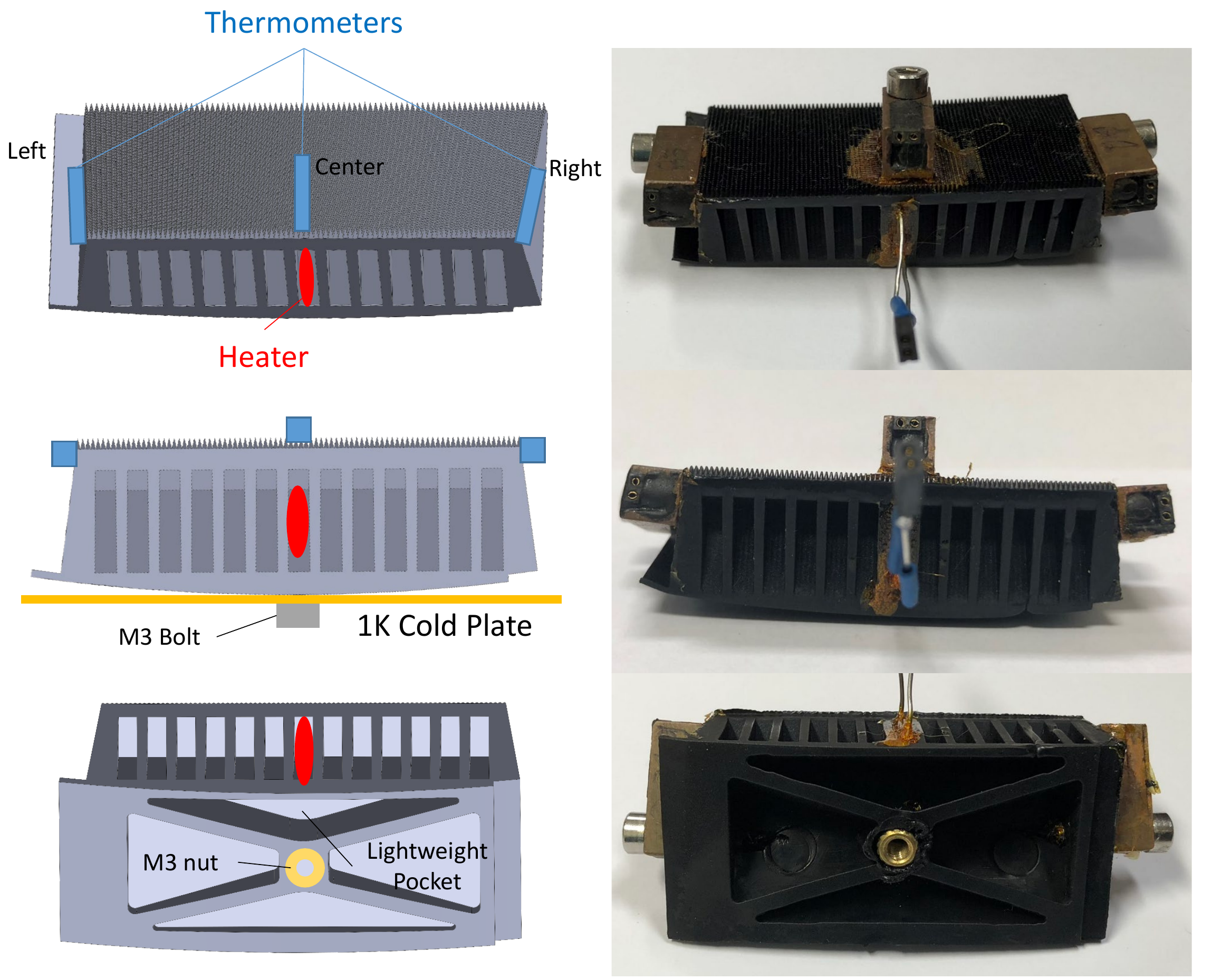}
    \caption{Thermal conductance test setup. Sketches on the left show the positions of the thermometers and the heater from three different perspectives; photos on the right show the actual setup from corresponding perspectives. The top row shows a top view of the positions of three thermometers and a heater. The middle row shows how the tile is mounted to the 1\,K cold plate with a stainless steel M3 bolt. The temperature of the 1\,K cold plate is continuously monitored by a fourth thermometer. Given the curved-to-flat interface, gaps are clearly seen around the single screw contact. The bottom row shows the positions of the M3 nut and the lightweight pocket together with the position of the heater. Because of the upper pocket, power from the heater does not short to the M3 nut directly.}
    \label{fig:thermal_test_setup}
\end{figure}

The manufactured tiles (shown in Fig.~\ref{fig:real_tile}) weigh $\sim$\,7.7\,g each. We inspected the tip of the pyramids under a microscope. The tips are rounded with a $\sim$150\,$\mu$m radius, a quarter of the pyramid spacing. The actual height to pitch ratio reduced from the designed 2.5:1 to around 1.5:1. However, a square version of the MMA tiles were later manufactured with sharper tips, almost achieving the designed 2.5:1 ratio (Fig.~\ref{fig:square_tiles}).

\section{Thermal Testing}
\label{sec:thermal_testing}

For our current application in the front part of the SO LATR optics tubes, the tiles will be used in a 4\,K radiation environment and cooled down to 4\,K. The concern was that the 4\,K radiation load can potentially heat up the absorbing surface if the tile's thermal conductance is insufficient. In addition, other applications may require cooling the tile to 1\,K, for example on the Lyot stop of the LATR optics tubes and in the SO Small Aperture Telescopes (SO SAT Cryogenic Baffling Development, 2020 in preparation). Thus, our thermal conductance test is to show whether the tile can be cooled to 1\,K in a 4\,K radiation environment. In a 4\,K radiation environment, the estimated radiation loading per tile is $\sim$\,40\,nW. The test must therefore show that 40\,nW applied to the absorbing surface does not appreciably raise the temperature of the surface.

Initially, we tested mechanical integrity of several carbon-loaded plastic materials under cryogenic temperatures.\footnote{Tested materials include carbon loaded polypropylene, carbon loaded polyethylene, carbon-fiber loaded polypropylene, and carbon loaded polyurethane.} Samples were repeatedly immersed in liquid nitrogen to verify that they would survive multiple thermal cycles. Considering both the cryogenic survivability and the injection-molding manufacturability, carbon-loaded thermalplastic polyurethane (TPU) was chosen for initial production. After the tiles were manufactured, the thermal conductance was tested at 1\,K in a dilution-cooled cryostat. One tile was first prepared with a resistive heater and three ruthenium oxide (ROX) thermometers installed (see Fig.~\ref{fig:thermal_test_setup}). The heater was installed in a pocket on the back; the three thermometers were installed in three different positions on the absorbing surface. For mechanical and thermal conductance purposes, the heater was glued in one lightweighting pocket with GE Varnish.\footnote{GE Varnish (also known as IMI 7031, GE 7031 or VGE 7031)} The three thermometers were then bolted and glued (with GE Varnish) to the tile. The three thermometers were distributed in the center and on two sides of the absorbing surface. The absorbing surface, being far away from the mounting surface, is predicted to have the most significant temperature difference from the cold plate. 
The heater and the thermometers were each electrically connected to the outside of the cryostat via four cryogenic wires. All of the wires together introduced $<1$\,nW total thermal conductance, much less than the 40\,nW power level being tested.

\begin{figure}
    \centering
    \includegraphics[width = .45\textwidth]{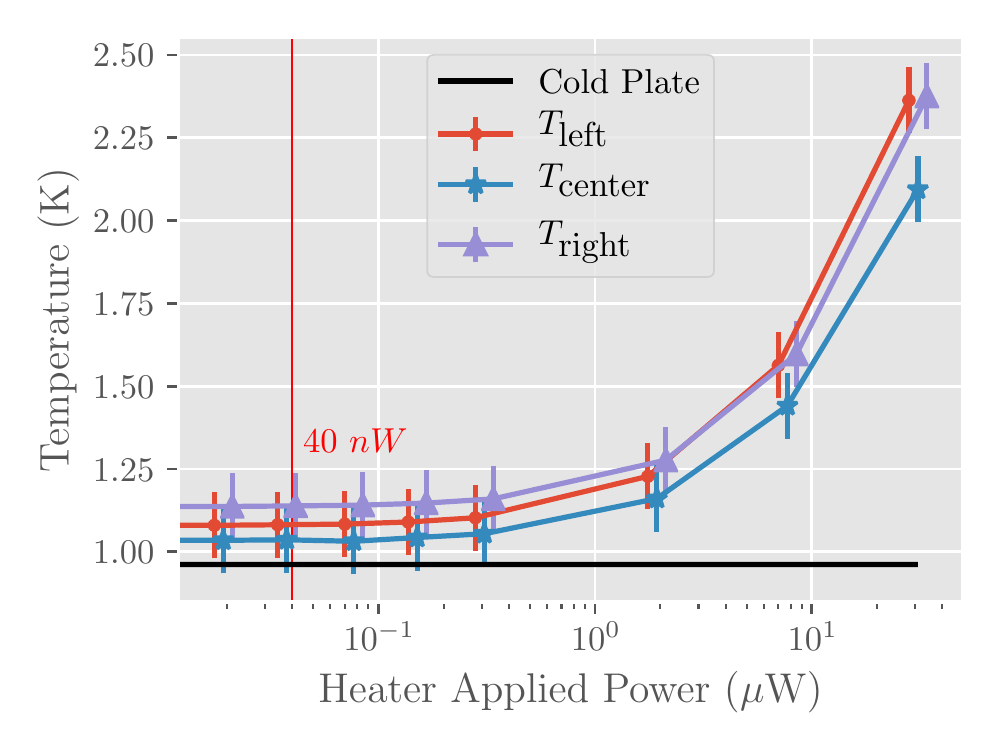}
    \caption{Thermal conductance measurement. The x-axis shows the additional power applied through the heater (in addition to the static 18\,nW from each thermometer); the y-axis shows the equilibrium temperatures for each of the three thermometers installed on the tile. The location of each thermometers are illustrated in Fig.~\ref{fig:thermal_test_setup}. The 0.1\,K uncertainty in the thermometer temperatures is expressed in error bars. The three sets of temperature data are shifted for clarity. The black line at the bottom records the temperature of the 1\,K cold plate. The vertical red line annotates the position of the 40\,nW radiation loading in a 4\,K radiation environment.}
    \label{fig:thermal_conductance}
\end{figure}

The prepared tile was fastened to the 1\,K plate within the cryostat using a stainless steel M3 bolt. The M3 bolt passed through the 1\,K cold plate and was threaded into the brass nut embedded in the tile. The M3 bolt was torqued to 0.35\,N$\cdot$m with a torque screw driver. The same value was used when installing the tiles in the SO LATR optics tubes. The radiation environment around the tile was also 1\,K. Since the tile has a curved bottom surface, the thermal contact with the flat cold plate was primarily at the exposed surface of the brass nut. The middle sketch in Fig.~\ref{fig:thermal_test_setup} illustrates the curved-to-flat surface interface. The resistive heater had a resistance of 120\,k$\Omega$ at room temperature and was measured to have a resistance of 129.7\,k$\Omega$ at $\sim$\,1\,K. This resistance was much greater than the $\sim$1\,k$ \Omega$ wire resistance, meaning that the heat dissipation on the cables was negligible. The three thermometers installed on the tiles have an absolute accuracy of $\pm$\,0.1\,K at 1.4\,K; the thermometer installed on the 1\,K cold plate has an accuracy of $\pm$\,0.005\,K at 1\,K. During the measurement, the voltage applied across the heater was gradually increased with a one hour time interval between adjustments. This interval was much longer than the previously-measured 200-second system thermal relaxation time. The applied voltage was stepped up eight times, covering a power range of 20\,nW to 30\,$\mu$W. The three thermometers were read out at a rate of 5\,Hz during the entire measurement. The 1\,K cold plate temperature changed <\,0.001\,K throughout the test.

\begin{figure}
    \centering
    \includegraphics[width = .45\textwidth]{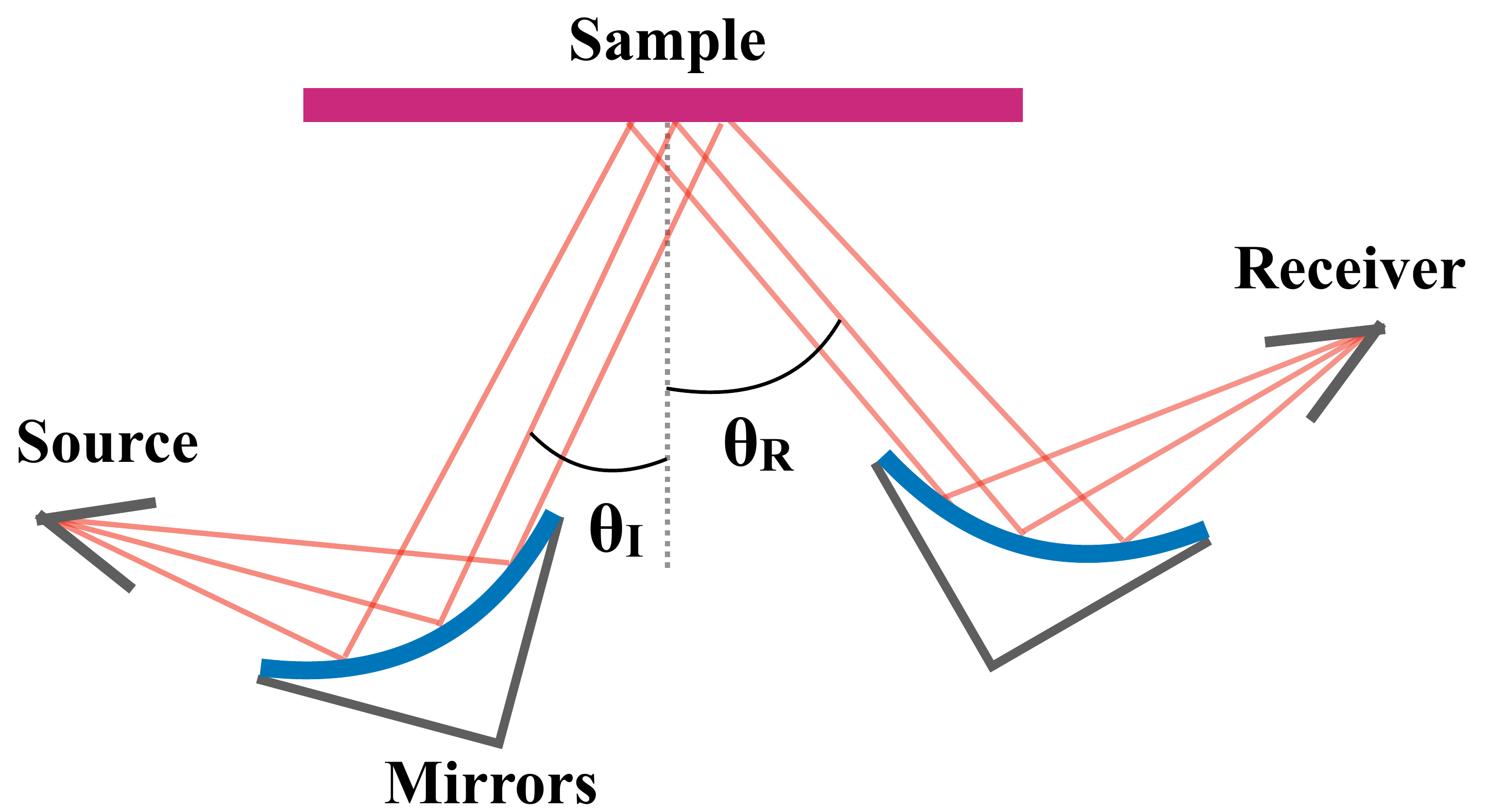}
    \caption{Optical measurement setup sketch. The orange lines represent the beam as it travels from the source feed-horn, collimated off of the first mirror and to the sample. After being reflected/scattered by the second mirror, the beam is focused into the receiver to be measured. Incident ($\theta_I$) and receiver ($\theta_R$) angles are controlled by a motor for measuring different angles of incidence and scattered power.}
    \label{fig:optical_setup}
\end{figure}

At each power level, three equilibrium temperatures were obtained from the three thermometers (see Fig.~\ref{fig:thermal_conductance}). With no heater power applied, all three thermometers exhibited a higher temperature than the cold plate. The center thermometer measured around 0.1\,K above the cold plate temperature, and the right and left side thermometers measured about 0.1\,K above the center thermometer. This can be explained by a combination of the absolute accuracy of the thermometers ($\pm$\,0.1\,K) and the thermometer self-heating. The self-heating comes from the current running through the thermometers generating around 18\,nW for each thermometer. However, the accuracy of the base temperatures did not affect the following measurement since we were interested in how the temperatures would change when extra loading was applied. The thermometers may not have an accurate absolute calibration, but they are sensitive to relative changes. 

\begin{figure*}
    \centering
    \includegraphics[width = .7\textwidth]{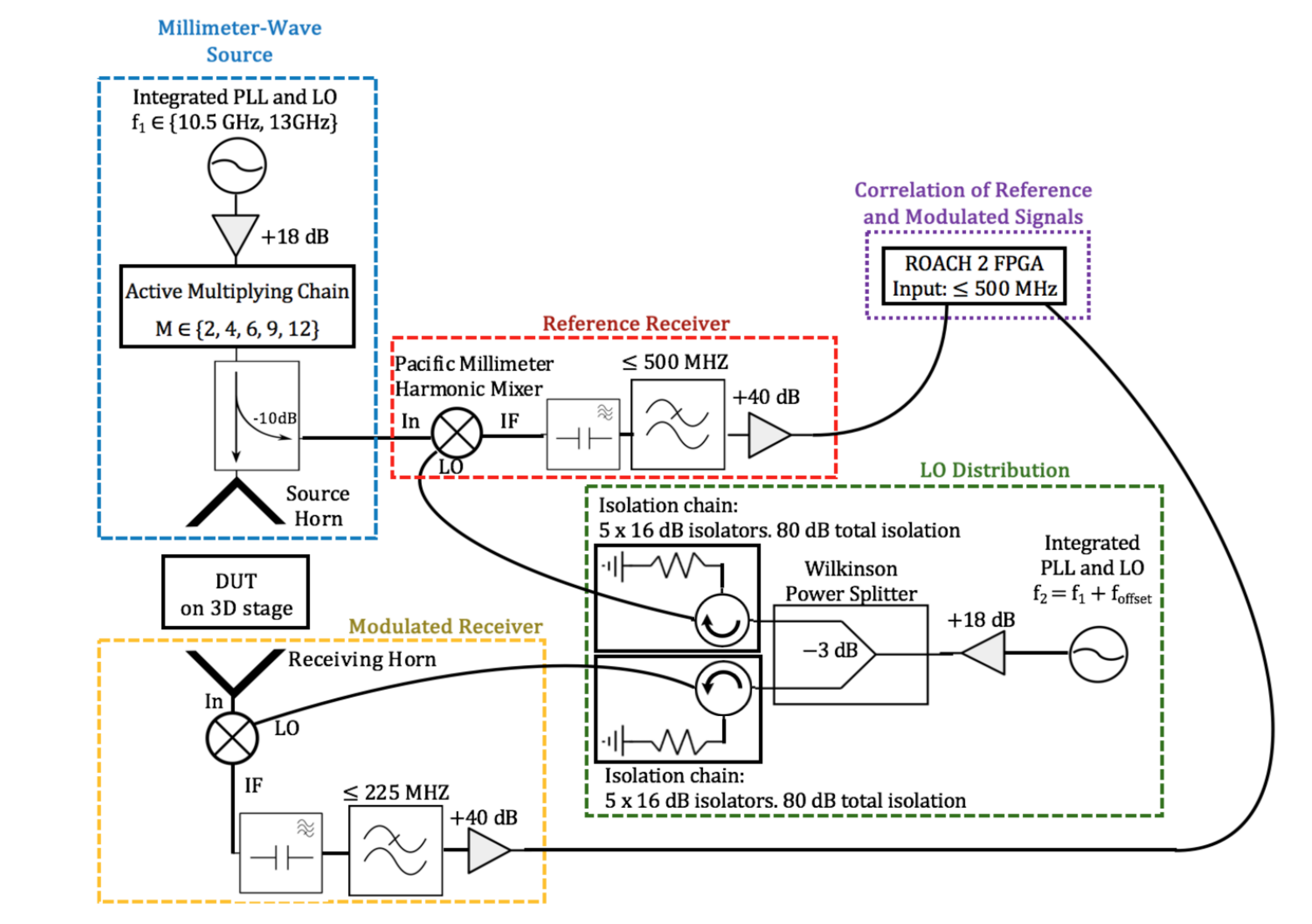}
    \caption{The correlation receiver setup diagram. A splitter sends one signal from a millimeter-wave source to a Pacific Millimeter harmonic mixer, while sending the same signal to be modulated by the device under test. The reference and modulated signal are mixed with an LO with an offset frequency $f_{\textrm{offset}}$ from the millimeter wave source. The two Pacific Millimeter harmonic mixers extract interference information caused by $f_{\textrm{offset}}$ and sends this to the ROACH-2 field programmable gate array (FPGA) board where the two signals are correlated.}
    \label{fig:roach}
\end{figure*}

\begin{figure}
    \centering
\includegraphics[width = .45\textwidth]{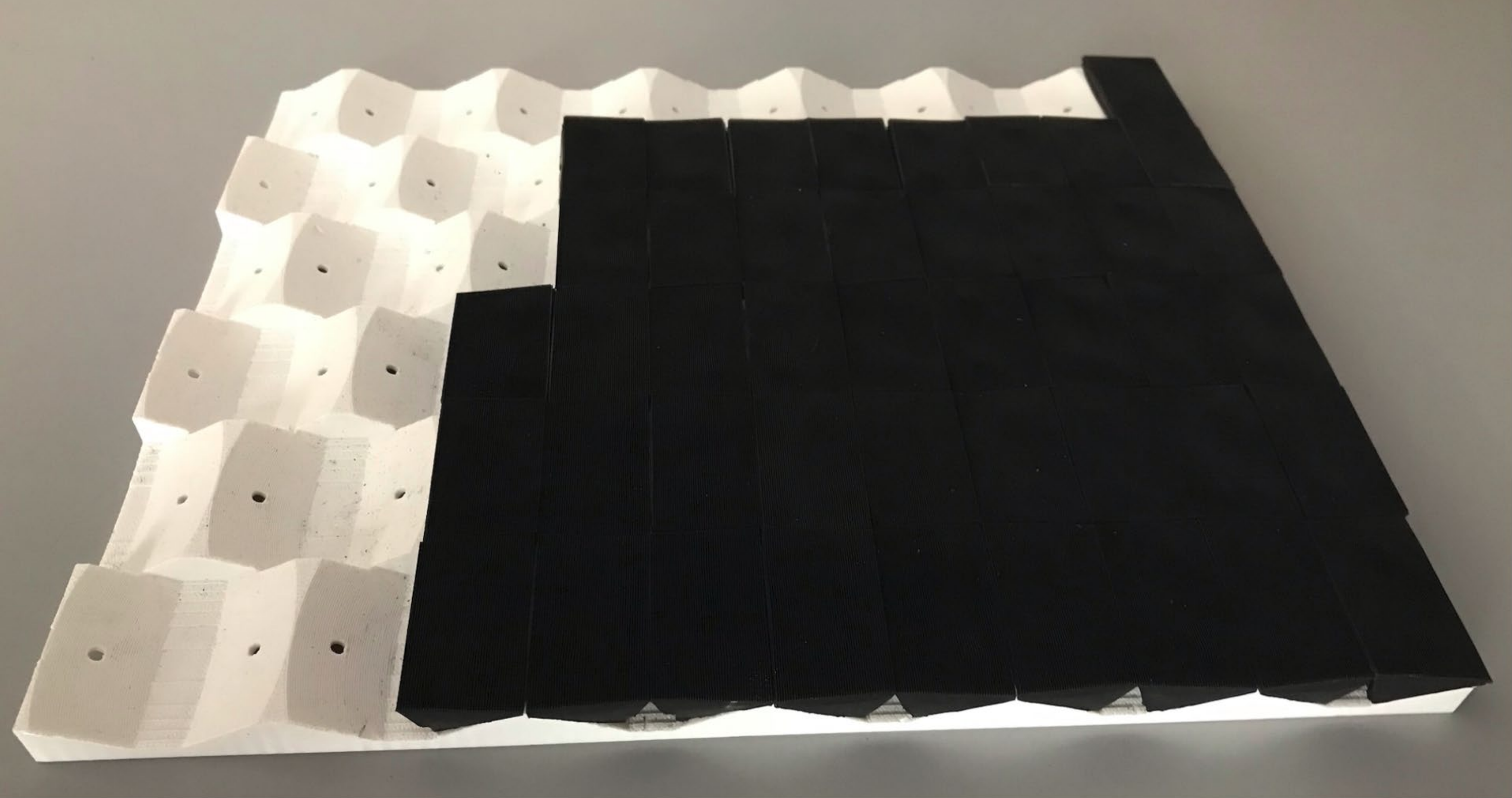}
    \caption{MMA measurement sample. MMA tiles are individually bolted into a 3D-printed plate such that each absorbing surface sits flat to form a 30.5$\times$30.5\,cm testing surface.}
    \label{fig:front_tiles}
\end{figure}

As illustrated in Fig.~\ref{fig:thermal_conductance}, no change in the absorbing surface temperature was measured below 100\,nW of applied extra power. As the applied loading reached 30\,$\mu$W, the temperatures from the three thermometers increased by $\sim$1\,K. A red vertical line is plotted in Fig.~\ref{fig:thermal_conductance} to emphasize the value of 40\,nW, which is the expected radiative loading from a 4\,K radiation environment. This result implies that a 4\,K radiation environment does not raise the temperature of the optical surface at 1\,K. Furthermore, Fig.~\ref{fig:thermal_conductance} shows that no significant temperature rises occur until around 300\,nW, implying that the tile can stay at 1\,K in a radiation environment as high as:
\begin{equation}
    T_{\textrm{upper}} = \left( \frac{P_{\textrm{upper}}}{P_{\textrm{4K}}}\right)^{\frac{1}{4}}\,T_\textrm{{4K}} =\left(\frac{300}{40}\right)^{\frac{1}{4}}\,4\,\textrm{K} \simeq 6.6\,\textrm{K}.
\end{equation}
As the final means of verification, the optics tube was successfully cooled to 4\,K in the SO LATR with 240 of these tiles installed, as shown in Fig.~\ref{fig:real_tile} on the right.

\section{Optical Testing}
\label{sec:optical_testing}
Optical properties of the MMA tiles were measured for diffuse reflection, or scattering, at 110\,GHz and specular reflection in the frequency range from 90\,GHz to 170\,GHz. The measurements also covered different angles of incidence. The results verified that the tiles achieved the designed high absorptivity and low reflection and scattering.

The optical measurements were conducted at room temperature. A consideration in using the technology at low temperatures is the following---the detailed composition and realization of the material (e.g., amorphous lamp black, activated carbon, pyrolytic carbon, graphite, etc.) influences the temperature dependence of the bulk resistivity and thus the dielectric function of the plastic \cite{Smith1956,Sihvola2008}. From our measurement of the bulk resistivity at 3\,K relative to ambient, this is a modest effect for the carbon loaded TPU formulation explored here. The design of the MMA tiles is insensitive to the dielectric function of the bulk dielectric mixture. Therefore, dramatic changes of the tile's optical properties are not expected in cryogenic temperatures down to 3\,K or lower. 

\begin{figure*}
    \centering
    \includegraphics[width = \textwidth]{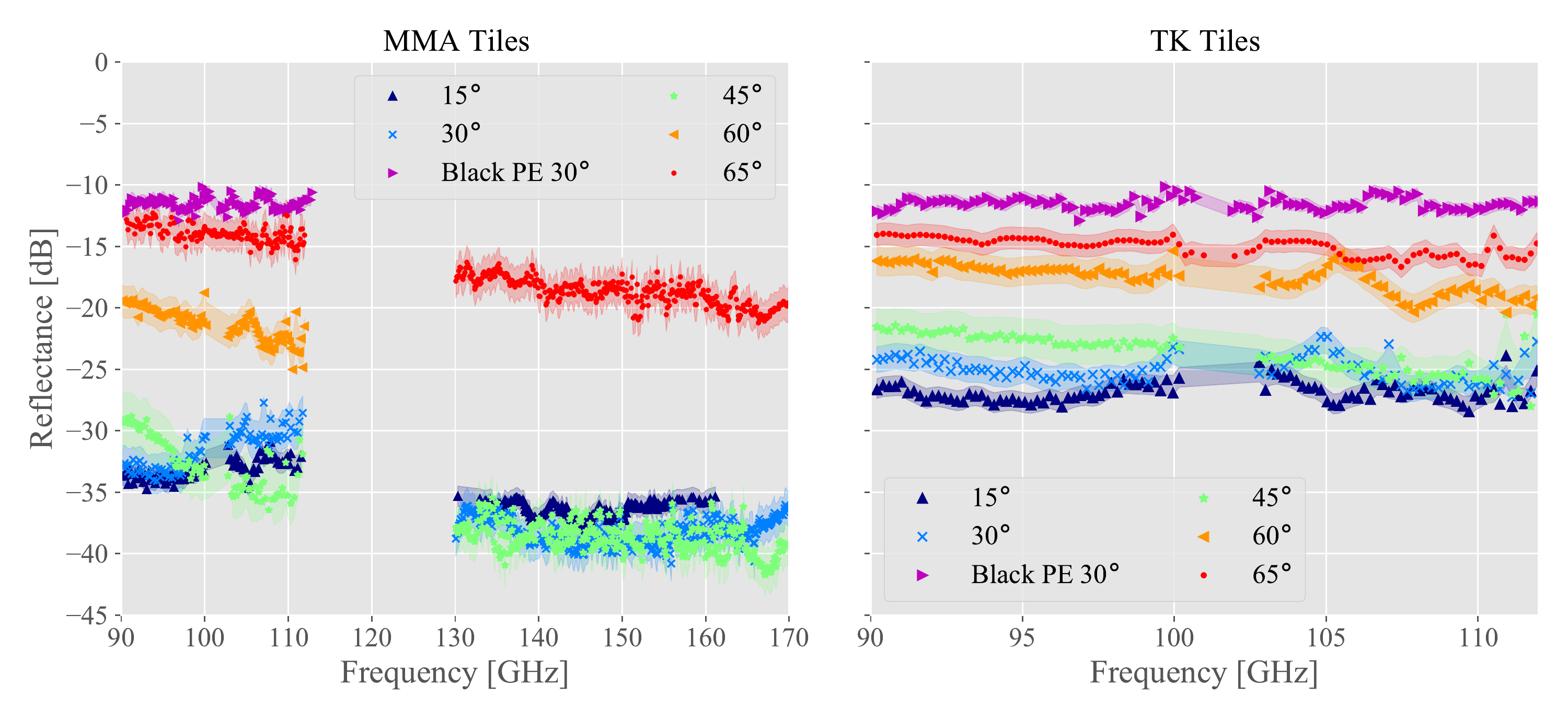}
    \caption{Specular reflectance measurements. The two panels show the specular reflectance measurements of the MMA tiles and the TK tiles at different angles of incidence, including $15^{\circ}$, $30^{\circ}$, $45^{\circ}$, $60^{\circ}$, and $65^{\circ}$. The samples are measured in W-band (90\,GHz to 110\,GHZ) and D-band (130\,GHz to 170\,GHz). The left panel shows the W-band (90\,GHz to 110\,GHZ) and D-band (130\,GHz to 170\,GHz) reflectance of the MMA tiles. The measurement gap in the left panel is due to the different frequency bands of the two sources used in the setup. The right panel measures the W-band (90\,GHz to 110\,GHZ) reflectance of the TK tiles. The two panels share the same y scale for comparison. The measurement of a flat black polyethylene sample without a AR-coating is included in both panels, which is conducted with a $30^{\circ}$ angle of incidence. Comparing the sample without an AR-coating (purple triangles) and our MMA tiles (blue crosses), $\sim$\,$-20$\,dB specular reflectance reduction is achieved with the metamaterial AR-coating.}
    \label{fig:reflection_result}
\end{figure*}

\subsection{Optical Hardware}

In our measurement setup, a source emits a millimeter-wave through a feed-horn toward a parabolic mirror, which then reflects a plane wave toward the sample, at an angle of incidence controlled by a rotary stage. The signal reflected off the sample propagates to a second mirror, followed by the receiver feed-horn~\cite{ches18}. The sketch of the setup is shown in Fig. \ref{fig:optical_setup}. Eccosorb HR-10 sheets\footnote{Emerson\,\&\,Cuming microwave products. Datasheet: \url{https://www.mouser.com/datasheet/2/987/HR-1591376.pdf}} are placed on surrounding surfaces to reject multipath propagation. The alignment of the receiver is controlled with a three-axis stage. The tilt of the sample is controlled with a three-point micrometer mount.

\subsection{Receiver Electronics}

A correlation receiver is used for these measurements. The receiver compares a reference tone to a signal which has passed through the optical path, creating an interference pattern between the two~\cite{ches18}. The correlation receiver is summarized in Fig.~\ref{fig:roach}.  The Re-configurable Open Architecture Computing Hardware (ROACH-2) board correlates the reference and modulated signals.\footnote{URL:  \url{https://casper.berkeley.edu/wiki/ROACH}}

The millimeter-wave source sends a signal, ranging from 10.5\,GHz to \,13 GHz, to a multiplier which passes through a passive multiplying chain. The W-band (90\,--\,110\,GHz) and D-band (130\,--\,180\,GHz) millimeter-wave sources use multiplication factors of 9 and 12, respectively. The signal is then modulated by the sample. The modulated and the reference signals are separately sent to two harmonic mixers. The mixers extract interference information caused by the offset frequency and send it to the correlation device. This receiver outputs the amplitude and phase of the signal in narrow (50 MHz) spectral bands.  Only the amplitude is used in this analysis, though we note that the phase information could be used to improve these measurements in the future.

\subsection{Measurement}

In order to measure their optical properties, an array of tiles are screwed down to a 3D-printed plate. The 3D-printed plate compensates the wedge shape of individual tiles so that absorbing surfaces of all tiles are aligned and oriented upward, forming a 30.5$\times$30.5\,cm flat surface (Fig.~\ref{fig:front_tiles}).

For comparison, we acquired the Tessellating TeraHertz RAM from TK Instruments.\footnote{Tessellating TeraHertz RAM. Product website:\url{http://www.terahertz.co.uk/tk-instruments/products/tesselatingterahertzram}} The TK tiles are 25\,mm $\times$ 25\,mm square tiles that can be tessellated to cover a flat surface. The pyramidal structures ($\sim$2.5\,mm in pitch and height) on the optical surface are designed to reduce the specular reflection and scattering in the 100\,--\,1000\,GHz region. 

Although the TK tiles and our MMA tiles both have pyramidal structures on the optical surface, they function differently. Pyramids in our MMA tiles, with the sub-wavelength scale, act more as a layer of medium with changing index of refraction; while pyramids in the TK tiles works more within the geometric optics regime to increase the number of bounces of the incoming radiation, due to the adopted surface geometry~\cite{Chuss2017}. The TK tiles have long been the preferred absorbers in millimeter wavelengths due to their optical performance. The measurement of the TK tiles ($\geq$\,200\,GHz) are available in the product website and published literature~\cite{Saily/etal:2004}. Because our MMA tiles are optimized to perform well at our operating wavelengths, we expect an improvement in performance. Even though our measurement only verifies the performance within 90\,--\,170\,GHz, our MMA tiles are designed to work from 90\,--\,270\,GHz following the same physical principles.

\subsubsection{Reflection Measurement}

Once the setup is aligned using an aluminum plate in place of the sample, a calibration data set is taken by measuring the specular reflectance of the same aluminum plate as a function of frequency. The sample measurements are then normalized using the aluminum plate data. 

Fig.~\ref{fig:reflection_result} shows the reflected power measured for the MMA and TK samples, with the MMA tiles measured from 90\,GHz to 170\,GHz and the TK tiles measured from 90\,GHz to 110\,GHz. The measured results demonstrate the relative performance of the two. For the MMA tiles, specular reflection is at sub-percent levels (<\,-20 dB) for all angles of incidence except for $65\degree$ ($\sim$\,-15\,dB). Note that $65\degree$ is higher than the angle of incidence upper limit in our application (see Section~\ref{sec:optical_design},\,\ref{sec:mechanical_design}). To demonstrate the effectiveness of the anti-reflection coating, a flat carbon-loaded polyethylene (PE) sample without a AR-coating was measured. During the test, a proper flat molding material sample was not available. Therefore, we measured the carbon-black loaded PE sample---with a lower dielectric function---as a lower limit of the reflectance from a flat molding material sample. With the coating, the MMA tiles reduce the specular reflection by $\sim$\,$-20$\,dB. Meanwhile, the specular reflection from the TK tiles is 5--10\,dB higher than that from the MMA tiles at different angles of incidence.

\subsubsection{Scattering Measurement}

\begin{figure}[t]
    \centering
    \includegraphics[width = .48\textwidth]{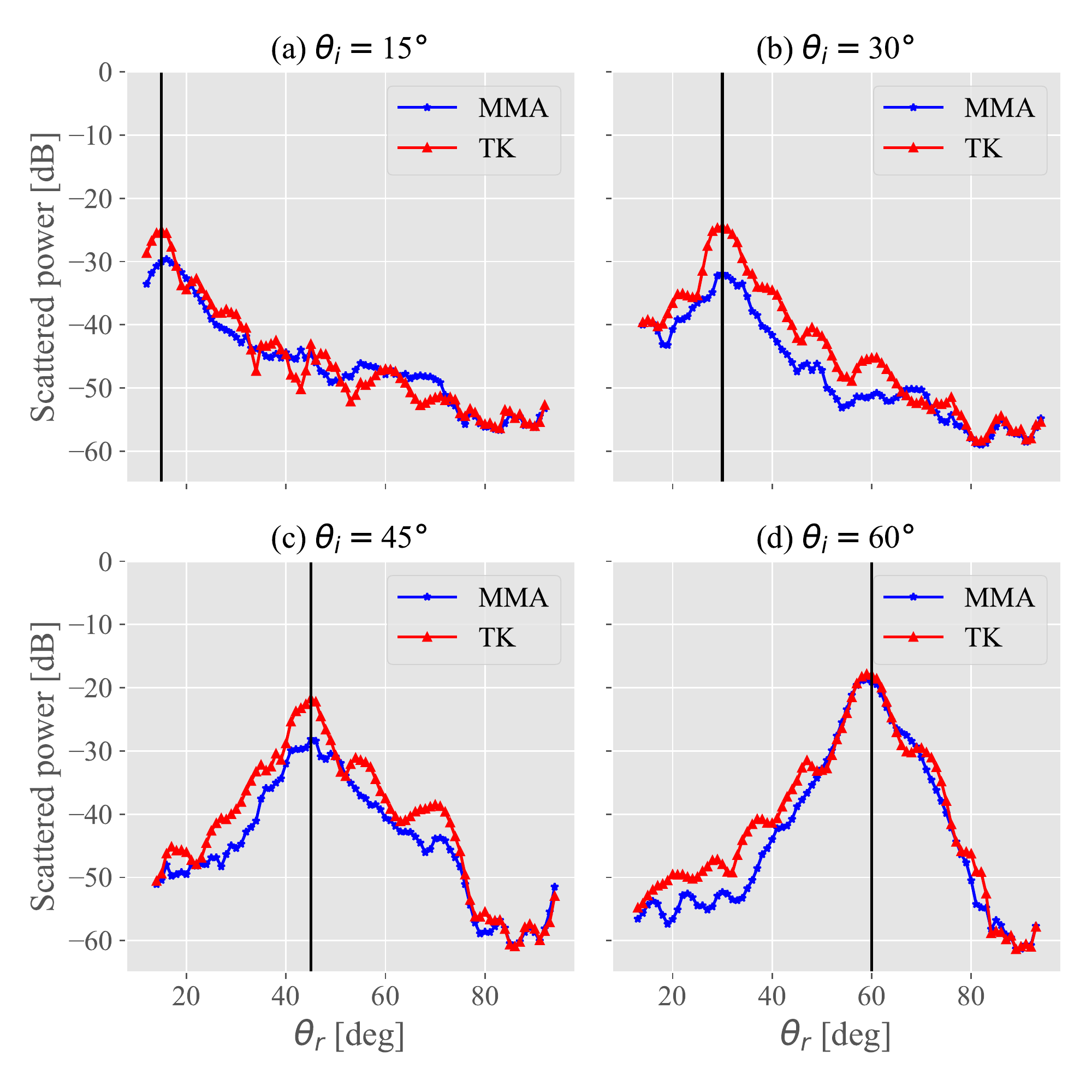}
    \caption{Scattering measurement. The four panels show front scattered power off the sample with the source located at four angles of incidence $\theta_i$: (a) \,15$\degree$, (b) \,30$\degree$, (c) \,45$\degree$, and (d) \,65$\degree$. The source was set at 110\,GHz. Each of the measured data points was averaged over 17 different sample rotation angles, from which the associated error bars were also estimated. The error bars are too small to be visible at these y scales. The vertical black lines indicate the position of the specular reflection directions. The blue data points show the results from the MMA tiles while the red data points show the results from the TK tiles. The TK tiles are measured to have an overall higher scattering than the MMA tiles, which leads to a higher integrated scattering power as shown in Table~\ref{tab:scat}.}
    \label{fig:scatter}
\end{figure}

The scattered power off the surface of the sample is determined by measuring the received power as a function of the receiver's angle, fixing the source feed horn at a set angle (Fig. \ref{fig:optical_setup}). The source (incident) and receiver (scattered) angles are controlled by two rotary stages. The radiative source sends one frequency and is held at a constant position (i.e. angle of incidence), while the receiver sweeps across different angles. Fig.~\ref{fig:scatter} shows the scattering measurement with a 110\,GHz source at four angles of incidence: 15$\degree$, 30$\degree$, 45$\degree$, and 60$\degree$. To account for the uneven surface of the samples, scattered power is measured and averaged over 17 rotations of the sample (Fig.~\ref{fig:front_tiles}). Then error bars (in Fig.~\ref{fig:scatter}) are estimated from the 17 measurements as the standard error of the mean. The measurements show that wide-angle scattering is suppressed to <\,-40\,dB (<\,0.01\,\%). All measurements are referenced to the specular reflection data from the aluminum plate. 

\begin{table}[h]
\centering
\begin{tabularx}{0.45\textwidth}{>{\centering\arraybackslash}X >{\centering\arraybackslash}X >{\centering\arraybackslash}X}
    \hline
    \hline
    $\theta_\text{I}$ & $15\degree$ & $30\degree$ \\
    \hline
    MMA & $0.49\pm0.03$\% &$0.12\pm0.002$\% \\
    TK  & $0.67\pm0.02$\% &$0.53\pm0.01$\%  \\
    \hline
    \hline
    $\theta_\text{I}$ & $45\degree$ & $60\degree$ \\
    \hline    
    MMA & $0.27\pm0.01$\% &$1.98\pm0.02$\% \\
    TK  & $0.83\pm0.01$\% &$2.24\pm0.02$\% \\
    \hline
\end{tabularx}
\caption{Integrated scattering power (in percentage) with different source angle of incidence $\theta_\text{I}$ at 110\,GHz for MMA and TK tiles. Integration is from $\theta_\text{R}=$ $0\degree$ to $45\degree$ away from the specular reflection direction. The cited error is calculated from the standard error of the mean over the 17 repeated measurements for each angle of incidence.}
  \label{tab:scat}
\end{table}

The integrated scattering power is calculated by integrating the scattering over the corresponding solid angles (Eq.~\ref{eq:scat}). The sample's fractional integrated scattering power, denoted as \textit{S}, is normalized by the total integrated input power where \textit{B} is the sample measurement, \textit{A} is the normalized aluminum plate reflection and scattering profile, and $\theta_r$ is receiver angle.

\begin{equation}
\text{S} =\int_{0}^{\frac{\pi}{4}} \text{B} \sin(\theta_\text{r}) d\theta_\text{r} \bigg/ \int_{0}^{\frac{\pi}{4}} \text{A} \sin(\theta_\text{r}) d\theta_\text{r}.
\label{eq:scat}
\end{equation}

We do not include the power beyond 45\dg{} ($\pi/4$) away from the specular reflection direction, considering its negligible contribution. We calculate the integrated scattering power for our MMA tiles and the TK tiles, listed in Table~\ref{tab:scat}. The results demonstrate the MMA tiles have lower integrated scattering power compared to the TK tiles, meeting the 1\% requirement for angle of incidence $\leq 45^{\circ}$.

\section{Future Applications}
\label{sec:future_applications}

\begin{figure}
    \centering
    \includegraphics[width=0.48\textwidth]{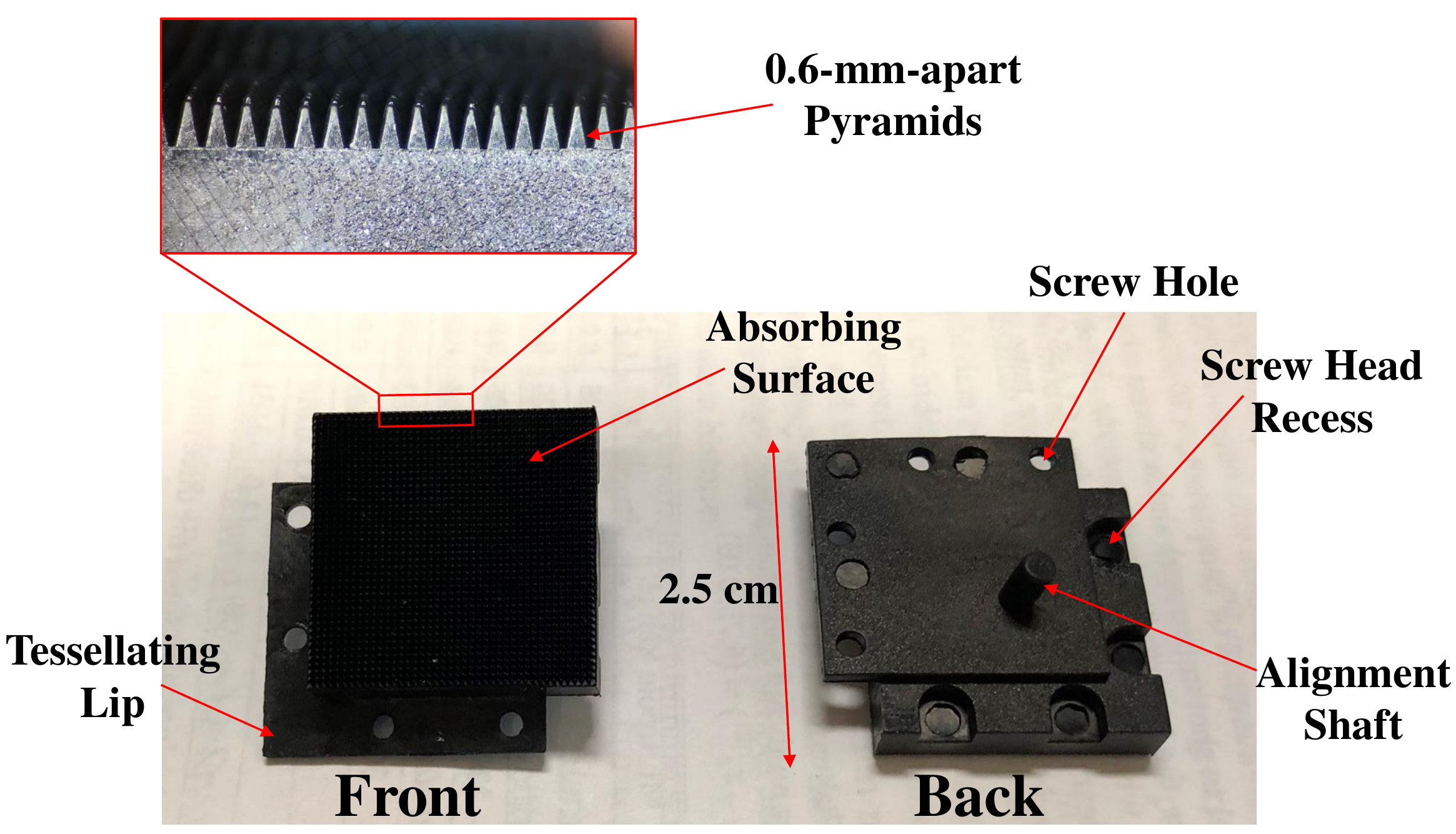}
    \caption{Manufactured square tiles. The 2.5\,cm $\times$ 2.5\,cm flat tiles are also manufactured via injection molding with the same pyramidal structure on the absorbing surface. The main photo on the bottom shows two square tiles from the front and the back. Major features on the tiles were denoted as well. A microscope photo of the pyramids is shown on the top left. The photo shows sharp features of the pyramids which are only 0.6-mm wide and 1.5-mm tall. The pyramids show sharper features compared to the ones in the tilted tiles (Fig.~\ref{fig:real_tile}). That is because the square tiles are smaller with a simpler shape, which facilitates the manufacturing process.}
    \label{fig:square_tiles}
\end{figure}

Flat and square MMA tiles were manufactured, in quantities of $> 10,000$, for more general use. The square tiles come in the size of 2.5\,cm $\times$ 2.5\,cm with the same anti-reflection coating (Fig.~\ref{fig:square_tiles}). Each of the flat tiles weigh $\sim$\,4\,g. Tessellating features cover the sides of the square tiles: two sides of the square overhang as lips while the other two sides indent. These features enable seamlessly tessellating copious number of tiles together. Given the limited thickness ($\sim$\,6\,mm), an injected nut cannot be installed; instead a shaft was designed on the back for alignment. For fastening, four over-sized M2 screw holes were designed on the tessellating lip. Correspondingly, four recesses for the screw head (from adjacent tiles) were designed on the tessellating dent (see Fig.~\ref{fig:square_tiles}). This version of MMA tile is designed to be applied on any flat surfaces by tessellation. They are already used to cover both sides of the Lyot stop in the SO LATR optics tubes (as shown in Fig.~\ref{fig:latr_ot}). The tile-covered Lyot stop assembly has been tested at 4\,K for thermal properties. 

Given the vast potential in customization, the MMA tiles provides a viable solution to blacken cryogenic surfaces with diverse geometries. Therefore, the technology can be useful for many future cryogenic experiments.

\section{Conclusion} 
The meta-material absorber (MMA) provides an effective, novel, and low-cost solution to absorb millimeter-wave radiation at cryogenic temperatures. Once the injection mold is manufactured, the tiles are mass-produced at low cost (< 1/10 of conventional methods) with a thousands-per-month production rate. Both the initial mold machining and injection molding were performed completely by an external shop, without any involvement from the research group. In addition, the technology also enables a high level of customization in the design. For future cryostat development, customized tiles can be designed and manufactured in this way to achieve the enhanced absorption properties, while fitting the geometry of specific cryostat designs.

The absorber was successfully cooled down to 1\,K with a single thermal contact. The optical properties, including specular reflection and scattering, were measured at different angles of incidence in the frequency band from 90\,GHz to 170\,GHz. The specular reflection is $< -30$\,dB for angles of incidence $\leq45$\dg{}. Even at 65\,\dg{}, the specular reflection is only $\sim-15$\,dB. The integrated scattering power is less than 1\% with the angle of incidence $\leq45$\dg{}. Even though this application was tailored for the frequency band from 90\,GHz to 270\,GHz, the metamaterial anti-reflection coating parameters can be suitably adjusted for use at longer wavelengths. The carbon grain size of this commercially available material and the injection molding process limit the suitability of direct scaling this design to shorter wavelengths. The use of a similar dielectric mixture with carbon lamp black and a lower volume filing fraction could be employed to address this challenge.

The design philosophy breaks the overall surface into basic tiles, facilitating their application to an extended surface. Unlike painted absorbing surfaces, the modularized design allows for the easy replacement of tiles, should part of the surface be damaged. The wedge-shape tile introduced in this paper is designed to work in a specific cylinder for Simons Observatory and CCAT-Prime. But different geometries, such as flat square tiles, are also available with similar optical performance.

\section{Funding Information}
This work was funded by the Simons Foundation (Award \#457687, B.K.), the University of Michigan, and the University of Pennsylvania. Zhilei Xu is supported by the Gordon and Betty Moore Foundation. Grace Chesmore is supported by the National Science Foundation (NSF) Graduate Student Research Fellowship (1650114). Frederick Matsuda acknowledges the support from the World Premier International Research Center Initiative (WPI), MEXT, Japan.

\section{Disclosures}
The authors declare no conflicts of interest.

\bibliography{bib_mma}

\begin{thebibliography}{10}
\newcommand{\enquote}[1]{``#1''}

\bibitem{gali18}
N.~Galitzki \emph{et~al.}, \enquote{{The Simons Observatory: instrument
  overview},} in \emph{Millimeter, Submillimeter, and Far-Infrared Detectors
  and Instrumentation for Astronomy IX,} , vol. 10708 J.~Zmuidzinas and J.-R.
  Gao, eds., International Society for Optics and Photonics (SPIE, 2018), pp. 1
  -- 13.

\bibitem{so19}
{Simons Observatory Collaboration} \emph{et~al.}, \enquote{The simons
  observatory: science goals and forecasts,} {\protect\JournalTitle{Journal of
  Cosmology and Astroparticle Physics}} \textbf{2019}, 056--056 (2019).

\bibitem{xu/etal:2020c}
Z.~Xu, T.~Bhandarkar, G.~Coppi, A.~Kofman, J.~Orlowski-Scherer, N.~Zhu
  \emph{et~al.}, \enquote{{The Simons Observatory: the Large Aperture Telescope
  Receiver (LATR) integration and validation results},} in \emph{Millimeter,
  Submillimeter, and Far-Infrared Detectors and Instrumentation for Astronomy
  X,} , vol. 11453 J.~Zmuidzinas and J.-R. Gao, eds., International Society for
  Optics and Photonics (SPIE, 2020).

\bibitem{zhu18}
N.~Zhu, J.~L. Orlowski-Scherer, Z.~Xu \emph{et~al.}, \enquote{{Simons
  Observatory large aperture telescope receiver design overview},} in
  \emph{Millimeter, Submillimeter, and Far-Infrared Detectors and
  Instrumentation for Astronomy IX,} , vol. 10708 J.~Zmuidzinas and J.-R. Gao,
  eds., International Society for Optics and Photonics (SPIE, 2018), pp. 259 --
  273.

\bibitem{orlo18}
J.~L. Orlowski-Scherer, N.~Zhu, Z.~Xu \emph{et~al.}, \enquote{{Simons
  Observatory large aperture receiver simulation overview},} in
  \emph{Millimeter, Submillimeter, and Far-Infrared Detectors and
  Instrumentation for Astronomy IX,} , vol. 10708 J.~Zmuidzinas and J.-R. Gao,
  eds., International Society for Optics and Photonics (SPIE, 2018), pp. 644 --
  657.

\bibitem{coppi/etal:2018}
G.~Coppi, Z.~Xu \emph{et~al.}, \enquote{{Cooldown strategies and transient
  thermal simulations for the Simons Observatory},} in \emph{Millimeter,
  Submillimeter, and Far-Infrared Detectors and Instrumentation for Astronomy
  IX,} , vol. 10708 J.~Zmuidzinas and J.-R. Gao, eds., International Society
  for Optics and Photonics (SPIE, 2018), pp. 246 -- 258.

\bibitem{ali20}
A.~M. {Ali} \emph{et~al.}, \enquote{{Small Aperture Telescopes for the Simons
  Observatory},} {\protect\JournalTitle{Journal of Low Temperature Physics}}
  \textbf{200}, 461--471 (2020).

\bibitem{iuliano/etal:2018}
J.~Iuliano, J.~Eimer, L.~Parker, G.~Rhoades \emph{et~al.}, \enquote{{The
  Cosmology Large Angular Scale Surveyor receiver design},} in
  \emph{Millimeter, Submillimeter, and Far-Infrared Detectors and
  Instrumentation for Astronomy IX,} , vol. 10708 J.~Zmuidzinas and J.-R. Gao,
  eds., International Society for Optics and Photonics (SPIE, 2018), pp. 259 --
  277.

\bibitem{thornton/etal:2016}
R.~J. Thornton \emph{et~al.}, \enquote{{THE} {ATACAMA} {COSMOLOGY} {TELESCOPE}:
  {THE} {POLARIZATION}-{SENSITIVE} {ACTPol} {INSTRUMENT},}
  {\protect\JournalTitle{The Astrophysical Journal Supplement Series}}
  \textbf{227}, 21 (2016).

\bibitem{sharp/etal:2008}
E.~H. {Sharp}, D.~J. {Benford}, D.~J. {Fixsen}, S.~F. {Maher}, C.~T. {Marx},
  J.~G. {Staguhn}, and E.~J. {Wollack}, \enquote{{Design and performance of a
  high-throughput cryogenic detector system},} in \emph{Millimeter and
  Submillimeter Detectors and Instrumentation for Astronomy IV,} , vol. 7020 of
  \emph{Society of Photo-Optical Instrumentation Engineers (SPIE) Conference
  Series} W.~D. {Duncan}, W.~S. {Holland}, S.~{Withington}, and
  J.~{Zmuidzinas}, eds. (2008), p. 70202L.

\bibitem{Wollack2008}
E.~J. Wollack, D.~J. Fixsen, R.~Henry, A.~Kogut, M.~Limon, and P.~Mirel,
  \enquote{Electromagnetic and thermal properties of a conductively loaded
  epoxy,} {\protect\JournalTitle{International Journal of Infrared and
  Millimeter Waves}} \textbf{29}, 51--61 (2008).

\bibitem{petroff/etal:2019}
M.~{Petroff}, J.~{Appel}, K.~{Rostem}, C.~L. {Bennett}, J.~{Eimer},
  T.~{Marriage}, J.~{Ramirez}, and E.~J. {Wollack}, \enquote{{A 3D-printed
  broadband millimeter wave absorber},} {\protect\JournalTitle{Review of
  Scientific Instruments}} \textbf{90}, 024701 (2019).

\bibitem{wollack/etal:2016}
E.~J. Wollack, A.~M. Datesman, C.~A. Jhabvala, K.~H. Miller, and M.~A. Quijada,
  \enquote{A broadband micro-machined far-infrared absorber,}
  {\protect\JournalTitle{Review of Scientific Instruments}} \textbf{87}, 054701
  (2016).

\bibitem{ding/etal:2012}
F.~Ding, Y.~Cui, X.~Ge, Y.~Jin, and S.~He, \enquote{Ultra-broadband microwave
  metamaterial absorber,} {\protect\JournalTitle{Applied Physics Letters}}
  \textbf{100}, 103506 (2012).

\bibitem{watts/liu/padilla:2012}
C.~M. Watts, X.~Liu, and W.~J. Padilla, \enquote{Metamaterial electromagnetic
  wave absorbers,} {\protect\JournalTitle{Advanced Materials}} \textbf{24},
  OP98--OP120 (2012).

\bibitem{vavgiakis/etal:2018}
E.~M. Vavagiakis \emph{et~al.}, \enquote{{Prime-Cam: a first-light instrument
  for the CCAT-prime telescope},} in \emph{Millimeter, Submillimeter, and
  Far-Infrared Detectors and Instrumentation for Astronomy IX,} , vol. 10708
  J.~Zmuidzinas and J.-R. Gao, eds., International Society for Optics and
  Photonics (SPIE, 2018), pp. 187 -- 202.

\bibitem{gudmundsson/etal:2020}
J.~E. {Gudmundsson}, P.~A. {Gallardo}, R.~{Puddu}, S.~R. {Dicker}
  \emph{et~al.}, \enquote{{The Simons Observatory: Modeling Optical Systematics
  in the Large Aperture Telescope},} {\protect\JournalTitle{arXiv e-prints}}
  arXiv:2009.10138 (2020).

\bibitem{hill/etal:2018}
C.~A. Hill, S.~M.~M. Bruno, S.~M. Simon \emph{et~al.}, \enquote{{BoloCalc: a
  sensitivity calculator for the design of Simons Observatory},} in
  \emph{Millimeter, Submillimeter, and Far-Infrared Detectors and
  Instrumentation for Astronomy IX,} , vol. 10708 J.~Zmuidzinas and J.-R. Gao,
  eds., International Society for Optics and Photonics (SPIE, 2018), pp. 698 --
  718.

\bibitem{Chuss2017}
D.~T. Chuss, K.~Rostem, E.~J. Wollack, L.~Berman, F.~Colazo, M.~DeGeorge,
  K.~Helson, and M.~Sagliocca, \enquote{A cryogenic thermal source for detector
  array characterization,} {\protect\JournalTitle{Review of Scientific
  Instruments}} \textbf{88}, 104501 (2017).

\bibitem{Smith1956}
A.~W. Smith and N.~S. Rasor, \enquote{Observed dependence of the
  low-temperature thermal and electrical conductivity of graphite on
  temperature, type, neutron irradiation, and bromination,}
  {\protect\JournalTitle{Phys. Rev.}} \textbf{104}, 885--891 (1956).

\bibitem{Sihvola2008}
A.~Sihvola, \emph{{Electromagnetic Mixing Formulas and Applications}}
  (Institution of Engineering and Technology, London, United Kingdom, 2008),
  {Electromagnetic Wave Series} ed.

\bibitem{ches18}
G.~E. {Chesmore}, T.~{Mroczkowski}, J.~{McMahon}, S.~{Sutariya}, A.~{Josaitis},
  and L.~{Jensen}, \enquote{{Reflectometry Measurements of the Loss Tangent in
  Silicon at Millimeter Wavelengths},} {\protect\JournalTitle{arXiv e-prints}}
  arXiv:1812.03785 (2018).

\bibitem{Saily/etal:2004}
J.~S{\"a}ily and A.~V. R{\"a}is{\"a}nen, \enquote{Characterization of
  submillimeter wave absorbers from 200--600 ghz,}
  {\protect\JournalTitle{International Journal of Infrared and Millimeter
  Waves}} \textbf{25}, 71--88 (2004).

\end{thebibliography}

\bibliographyfullrefs{bib_mma}

\end{document}